\documentclass[11pt]{article}
\usepackage{epstopdf}
\usepackage{color}
\usepackage{subfigure}
\usepackage{amsmath}
\usepackage{amssymb}
\usepackage{graphicx,color}
\usepackage{cite}
\usepackage{enumerate}
\usepackage{amsthm}
\usepackage{amsfonts,mathrsfs}
\usepackage{geometry}
\usepackage{ifthen}
\usepackage{graphicx}
\usepackage{float}
\usepackage{bm}
\usepackage[caption=false]{subfig}
\usepackage[utf8]{inputenc}

\begin{document}

\title{\bf Sum uncertainty relations based on $(\alpha,\beta,\gamma)$ weighted Wigner-Yanase-Dyson skew information}

\vskip0.1in
\author{\small Cong Xu$^1$, Zhaoqi Wu$^1$\thanks{Corresponding author. E-mail: wuzhaoqi\_conquer@163.com},
Shao-Ming Fei$^{2,3}$\\
{\small\it  1. Department of Mathematics, Nanchang University,
Nanchang 330031, P R China}\\
{\small\it  2. School of Mathematical Sciences, Capital Normal University, Beijing 100048, P R China}\\
{\small\it  3. Max-Planck-Institute for Mathematics in the Sciences,
04103 Leipzig, Germany} }

\date{}
\maketitle

\noindent {\bf Abstract} {\small }\\
We introduce ($\alpha,\beta,\gamma$) weighted Wigner-Yanase-Dyson
(($\alpha,\beta,\gamma$) WWYD) skew information and
($\alpha,\beta,\gamma$) modified weighted Wigner-Yanase-Dyson
(($\alpha,\beta,\gamma$) MWWYD) skew information. We explore the sum
uncertainty relations for arbitrary $N$ mutually noncommutative
observables based on ($\alpha,\beta,\gamma$) WWYD skew information.
A series of uncertainty inequalities are derived.
We show by detailed example that our results cover and improve the previous
ones based on the original Wigner-Yanase (WY) skew
information. Finally, we establish new sum uncertainty
relations in terms of the ($\alpha,\beta,\gamma$) MWWYD skew information
for arbitrary $N$ quantum channels.

\vskip0.2in

\noindent {\bf 1. Introduction}\\\hspace*{\fill}\\
As one of the most essential features of the quantum world, the uncertainty principle has been widespread concerned since Heisenberg \cite{HW} proposed the notions of uncertainties in measuring non-commuting observables. For arbitrary two observables $A$ and $B$, the well-known Heisenberg-Robertson \cite{RH} uncertainty relation with respect to a quantum state $|\psi\rangle$ says that,
\begin{equation}\label{eq1}
\Delta A\Delta B\geq \frac{1}{2}|\langle\psi|[A,B]|\psi\rangle|,
\end{equation}
where $[A,B]=AB-BA$ and
$\Delta\Omega=\sqrt{\langle\psi|{\Omega}^2|\psi\rangle-{\langle\psi|\Omega|\psi\rangle}^2}$
is the standard deviation of an observable $\Omega$. Many different characterizations and quantifications of quantum uncertainty have been proposed in terms of
entropy \cite{DD,MHU,WSWA,WSYS,RAE,CB1,RLPZ,KK,HMJW},
variance \cite{GUDDER,DD1,DD2,SL}, under successive
measurements \cite{SMD,DJPS,BKFT,ZJZY}, and with majorization
techniques \cite{PZRL,RLPZ,RL,FSGV}.

The quantum uncertainty can also be characterized by skew information.
The Wigner-Yanase (WY) information and Wigner-Yanase-Dyson (WYD) skew
information associated to a quantum state $\rho$ and an observable $A$ have
been defined in \cite{WY}. The WYD skew information has been further
extended to the generalized Wigner-Yanase-Dyson (GWYD) skew information
\cite{CL}. The relationship between WY skew information and the
uncertainty relation has been originally established by Luo and
Zhang \cite{LUO3}, and various types of uncertainty relations based
on the WY skew information, WYD skew information and GWYD skew
information have been presentd
\cite{LUO1,LUO2,MCF,LUO4,LUO5,LUO6,LUO7,LUO8,LUO9,LUO10,FURU2,FURU3,YANA1,YANA2,KO}.

By considering state-channel interaction, in \cite{SLYS} Luo and Sun
defined a quantity $\mathrm{I}_{\rho}(\Phi)$ and
its dual one $\mathrm{J}_{\rho}(\Phi)$, and explored the
complementarity relation between them. Wu, Zhang and Fei
introduced the non-Hermitian extension of the GWYD
skew information and generalized the complementarity relation to a
more general case in \cite{WU1,WU2}.

On the other hand, another generalization of the WYD skew information,
the weighted Wigner-Yanase-Dyson (WWYD) skew information, has been
introduced in \cite{FURU1}. As its non-Hermitian extension,
the modified weighted Wigner-Yanase-Dyson (MWWYD) skew information
has been defined and investigated in \cite{CZL}. Recently, by using the
convex combination, instead of the arithmetic mean of $\rho^\alpha$
and $\rho^{1-\alpha}$, the two-parameter extension of the Wigner-Yanase
skew information has been formulated \cite{Zhang}.

Recently, the sum uncertainty relations based on the variance and WY skew
information have attracted considerable
attention \cite{CB2,CB3,ZL,ZQF}. In \cite{CB2,CB3} Chen and Fei
proposed some uncertainty inequalities in terms of the sum of
variances, standard deviations and the WY skew information for arbitrary
$N$ mutually noncommutative observables, respectively. After that,
Zhang, Gao and Yan \cite{ZL} established a tighter uncertainty
relation via WY skew information for arbitrary $N$ mutually
noncommutative observables, which extend the results in \cite{CB3}.
Zhang and Fei \cite{ZQF} further improved the results in
\cite{ZL} and proposed new tighter bounds than the existing
ones. Cai \cite{CAL} generalized the sum uncertainty relations
for WY skew information introduced in \cite{CB3} to an arbitrary
metric-adjusted skew information version. Ren, Li, Ye and
Li \cite{RLYL} proposed tighter sum uncertainty relations than
the ones in \cite{CAL}.

In \cite{FSS} Fu, Sun and Luo established the uncertainty relations for two quantum channels based on the WY skew information for arbitrary operators. Afterwards, Zhang, Gao and Yan \cite{ZL} generalized the uncertainty relations for two quantum channels to arbitrary $N$ quantum channels. Zhang, Wu and Fei \cite{ZWF} further generalized the results in \cite{ZL} and proposed new bounds which are tighter than the existing ones. Cai \cite{CAL} confirmed that the results in \cite{FSS} also hold for all metric-adjusted skew information.

The remainder of this paper is structured as follows. In Section 2,
we recall some basic concepts and propose the definitions of
$(\alpha,\beta,\gamma)$ weighted Wigner-Yanase-Dyson
($(\alpha,\beta,\gamma)$ WWYD) skew information and
$(\alpha,\beta,\gamma)$ modified weighted Wigner-Yanase-Dyson
($(\alpha,\beta, \gamma)$ MWWYD) skew information. In Section 3, we
present uncertainty inequalities for arbitrary $N$ mutually
noncommutative observables in terms of the $(\alpha,\beta,\gamma)$
WWYD skew information. Especially, we show that when
$\alpha=\beta=\frac{1}{2}$, i.e., the $(\alpha,\beta,\gamma)$ WWYD
skew information reduce to the WY skew information, the lower
bounds of our inequalities improve the existing ones by a detailed
example. In Section 4, we explore the
$(\alpha,\beta,\gamma)$ MWWYD skew information-based sum uncertainty
relations for quantum channels. Some concluding remarks are given in
Section 5.

\vskip0.1in

\noindent {\bf 2. $(\alpha,\beta,\gamma)$ WWYD skew information and
$(\alpha,\beta,\gamma)$ MWWYD skew information}\\\hspace*{\fill}\\
Let $\mathcal{H}$ be a $d$-dimensional Hilbert space. Denote by
$\mathcal{B(H)}$, $\mathcal{S(H)}$ and $\mathcal{D(H)}$ the set of
all bounded linear operators, Hermitian operators and density
operators (positive operators with trace 1) on $\mathcal{H}$,
respectively. Mathematically, a quantum state and a quantum channel
are represented by a density operator and a completely positive
trace-preserving map, respectively.

For a quantum state $\rho\in \mathcal{D(H)}$ and an observable $A\in
\mathcal{S(H)}$, the {\it Wigner-Yanase} (WY) skew information
\cite{WY} is defined by
\begin{equation}\label{eq2}
\mathrm{I}_{\rho}(A):=-\frac{1}{2}\mathrm{Tr}\left([\sqrt{\rho},A]^2\right)
=\frac{1}{2}\|[\sqrt{\rho},A]\|^{2},
\end{equation}
where $\|\cdot\|$ denotes the Hilbert Schmidt norm,
$\|T\|=\sqrt{\mathrm{Tr}T^{\dag}T}$. $\mathrm{I}_{\rho}(A)$ is generalized by Dyson to
\begin{equation}\label{eq3}
\mathrm{I}_{\rho}^{\alpha}(A):=-\frac{1}{2}\mathrm{Tr}([\rho^{\alpha},A][\rho^{1-\alpha},A]),
\,\,~0\leq \alpha \leq 1,
\end{equation}
which is now called the {\it Wigner-Yanase-Dyson} (WYD) skew
information\cite{WY}. $\mathrm{I}_{\rho}^{\alpha}(A)$ is further generalized to \cite{CL}
\begin{equation}\label{eq4}
\mathrm{I}_{\rho}^{\alpha,\beta}(A)=-\frac{1}{2}\mathrm{Tr}([\rho^\alpha,A][\rho^\beta,
A]\rho^{1-\alpha-\beta}),~~~\alpha,\beta\geq 0,~\alpha+\beta\leq 1,
\end{equation}
which is termed as {\it generalized Wigner-Yanase-Dyson} (GWYD) skew
information.

Another generalization of the WYD skew information is given in \cite{FURU1},
\begin{align}\label{eq5}
\mathrm{K}_{\rho}^{\alpha}(A)=-\frac{1}{2}\mathrm{Tr}
\left(\left[\frac{\rho^\alpha+\rho^{1-\alpha}}{2},A\right]^2\right)
=\frac{1}{2}\left\|\left[\frac{\rho^\alpha+\rho^{1-\alpha}}{2},
A\right]\right\|^{2} ~,\,\,0\leq \alpha \leq 1,
\end{align}
which is called the {\it weighted Wigner-Yanase-Dyson} (WWYD) skew
information. The authors in \cite{CZL} proposed the {\it modified weighted
Wigner-Yanase-Dyson} (MWWYD) skew information, which
is the non-Hermitian extension of the WWYD skew information.

By replacing the arithmetic mean of $\rho^\alpha$ and $\rho^{1-\alpha}$
with their convex combination, the {\it two-parameter extension of
Wigner-Yanase skew information} has been introduced,
\begin{align}\label{eq6}
\mathrm{K}_{\rho,\gamma}^{\alpha}(A)\notag
=&-\frac{1}{2}\mathrm{Tr}\left([(1-\gamma)\rho^\alpha+\gamma\rho^{1-\alpha},A]^{2}\right)\\
=&\frac{1}{2}\left\|\left[(1-\gamma)\rho^\alpha+\gamma\rho^{1-\alpha},
A\right]\right\|^{2} ~,~\,\,0\leq \alpha \leq 1~,\,\,0\leq \gamma \leq 1.
\end{align}
For convenience, we call it the {\it $(\alpha,\gamma)$ weighted
Wigner-Yanase-Dyson} ($(\alpha,\gamma)$ WWYD) skew information in
this paper. Note that Eq. (\ref{eq6}) reduces to Eq. (\ref{eq5}) and
Eq. (\ref{eq2}) when $\gamma=\frac{1}{2}$ and $\alpha=\frac{1}{2}$,
respectively.

For a quantum state $\rho\in \mathcal{D(H)}$ and an observable $A\in
\mathcal{S(H)}$, we define the {\it $(\alpha,\beta,\gamma)$ weighted
Wigner-Yanase-Dyson} ($(\alpha,\beta,\gamma)$ WWYD) skew information
as
\begin{align}\label{eq7}
\mathrm{K}_{\rho,\gamma}^{\alpha,\beta}(A)\notag
=&-\frac{1}{2}\mathrm{Tr}([(1-\gamma)\rho^{\alpha}+\gamma\rho^{\beta},A]^{2}
\rho^{1-\alpha-\beta})\\
=&\frac{1}{2}\left\|\rho^\frac{1-\alpha-\beta}{2}\left[(1-\gamma)\rho^\alpha+\gamma\rho^\beta,
A\right]\right\|^{2},~~\alpha,\beta\geq 0,~\alpha+\beta\leq 1,0\leq
\gamma \leq 1.
\end{align}
Note that Eq. (\ref{eq7}) reduces to Eq. (\ref{eq6}) when
$\beta=1-\alpha$.

We also define the {\it $(\alpha,\beta,\gamma)$ modified weighted
Wigner-Yanase-Dyson} ($(\alpha,\beta,\gamma)$ MW\\WYD) skew
information with respect to a quantum state $\rho\in \mathcal{D(H)}$ and an
arbitrary operator $E\in \mathcal{B(H)}$ (not necessarily Hermitian),
\begin{align}\label{eq8}
\mathrm{K}_{\rho,\gamma}^{\alpha,\beta}(E)\notag
=&-\frac{1}{2}\mathrm{Tr}([(1-\gamma)\rho^{\alpha}+\gamma\rho^{\beta},E^{\dag}][(1-\gamma)\rho^{\alpha}+\gamma\rho^{\beta},E]\rho^{1-\alpha-\beta})\\
=&\frac{1}{2}\left\|\left[(1-\gamma)\rho^\alpha+\gamma\rho^\beta,
E\right]\rho^\frac{1-\alpha-\beta}{2}\right\|^{2},~~~\alpha,\beta\geq 0,~\alpha+\beta\leq 1,0\leq\gamma \leq 1,
\end{align}
which is the non-Hermitian extension of the $(\alpha,\beta,\gamma)$ WWYD
skew information. Note that Eq. (\ref{eq8}) reduces to Eq.
(\ref{eq10}) in \cite{WU2} when $\gamma=\frac{1}{2}$.

Following the idea in \cite{SLYS}, we further define the
$(\alpha,\beta,\gamma)$ MWWYD skew information of $\rho$ with
respect to a channel $\Phi$ as
\begin{equation}\label{eq9}
\mathrm{K}_{\rho,\gamma}^{\alpha,\beta}(\Phi)=\sum_{i=1}^{n}\mathrm{K}_{\rho,\gamma}^{\alpha,\beta}(E_i),
\end{equation}
where $\alpha,\beta\geq0,~\alpha+\beta\leq 1,0\leq \gamma \leq 1$,
and $E_i(i=1,2,\cdots,n)$ are Kraus operators of the channel $\Phi$,
i.e., $\Phi(\rho)=\sum_{i=1}^{n}E_i\rho E_i^{\dag}$.

\vskip0.1in

\noindent {\bf 3. Sum uncertainty relations for arbitrary $N$ mutually noncommutative observables in terms of the $(\alpha,\beta,\gamma)$ WWYD skew information}\\\hspace*{\fill}\\
We now provide several sum uncertainty relations in terms of the $(\alpha,\beta,\gamma)$ WWYD skew information for arbitrary $N$ mutually noncommutative observables.\\\hspace*{\fill}\\
{\bf Theorem 1} For arbitrary $N$ mutually noncommutative
observables $A_1, A_2,\cdots,A_N$ ($N\geq2$), we have
\begin{equation}\label{eq10}
\sum_{i=1}^{N}{\mathrm{K}_{\rho,\gamma}^{\alpha,\beta}(A_i)}
\geq \mathrm{max}\frac{1}{2(N-1)}\left\{\sum_{1\leq i<j\leq N} \mathrm{K}_{\rho,\gamma}^{\alpha,\beta}(A_i+A_j),\sum_{1\leq i<j\leq N} \mathrm{K}_{\rho,\gamma}^{\alpha,\beta}(A_i-A_j)\right\},
\end{equation}
where $\alpha,\beta\geq 0,~\alpha+\beta\leq 1,~0\leq \gamma \leq
1$.\\\hspace*{\fill}\\
\textit{Proof} By using the following equality,
\begin{align*}
2(N-1)\sum_{i=1}^{N} \| u_i\|^2=\sum_{1\leq i<j\leq N} \| u_i+u_j\|^2+\sum_{1\leq i<j\leq N} \| u_i-u_j\|^2,
\end{align*}
we have
\begin{align*}
\sum_{i=1}^{N} \| u_i\|^2\geq\frac{1}{2(N-1)}\sum_{1\leq i<j\leq N} \| u_i+u_j\|^2,
\end{align*}
and
\begin{align*}
\sum_{i=1}^{N} \| u_i\|^2\geq\frac{1}{2(N-1)}\sum_{1\leq i<j\leq N} \| u_i-u_j\|^2.
\end{align*}
Therefore,
\begin{align*}
\sum_{i=1}^{N} \| u_i\|^2\geq \mathrm{max} \frac{1}{2(N-1)}\left\{\sum_{1\leq i<j\leq N} \| u_i+u_j\|^2, \sum_{1\leq i<j\leq N} \| u_i-u_j\|^2\right\}.
\end{align*}
Setting
$u_i=\rho^\frac{1-\alpha-\beta}{2}\left[(1-\gamma)\rho^\alpha+\gamma\rho^\beta,A_i\right]$
and
$u_j=\rho^\frac{1-\alpha-\beta}{2}\left[(1-\gamma)\rho^\alpha+\gamma\rho^\beta,
A_j\right]$,
we get (\ref{eq10}). $\Box$

In particular, for $N=2$ from Theorem 1 we have the following Corollary 1.\\\hspace*{\fill}\\
{\bf Corollary 1} For arbitrary two noncommutative observables $A$ and $B$, we have
{\begin{equation}\label{eq11}
\mathrm{K}_{\rho,\gamma}^{\alpha,\beta}(A)+\mathrm{K}_{\rho,\gamma}^{\alpha,\beta}(B)\geq\mathrm{max}\frac{1}{2}\{\mathrm{K}_{\rho,\gamma}^{\alpha,\beta}(A+B),\mathrm{K}_{\rho,\gamma}^{\alpha,\beta}(A-B)\},
\end{equation}where $\alpha,\beta\geq 0,~\alpha+\beta\leq 1,~0\leq \gamma \leq 1$.

Note that (\ref{eq11}) of Corollary 1 reduces to the formula (\ref{eq3}) of Theorem 1 in \cite{CB3} when $\alpha=\beta=\frac{1}{2}$.
\\\hspace*{\fill}\\
{\bf Theorem 2} For arbitrary $N$ mutually noncommutative
observables $A_1, A_2,\cdots,A_N$ ($N\geq2$), we have
\begin{equation}\label{eq12}
\sum_{i=1}^{N}\sqrt{\mathrm{K}_{\rho,\gamma}^{\alpha,\beta}(A_{i})}
\geq\sqrt{\mathrm{K}_{\rho,\gamma}^{\alpha,\beta}\left(\sum_{i=1}^{N} A_i\right)}~,\,\,\alpha,\beta\geq 0,~\alpha+\beta\leq 1,~0\leq \gamma \leq 1,
\end{equation}
and
\begin{equation}\label{eq13}
\sum_{i=1}^{N}\sqrt{\mathrm{K}_{\rho,\gamma}^{\alpha,\beta}(A_{i})}
\geq\sqrt{\mathrm{K}_{\rho,\gamma}^{\alpha,\beta}\left(\sum_{i=1}^{N-1} A_i-A_N\right)}~,\,\,\alpha,\beta\geq 0,~\alpha+\beta\leq 1,~0\leq \gamma \leq 1.
\end{equation}
\textit{Proof} By using the norm inequality, we obtain
\begin{align*}
\sqrt{\mathrm{K}_{\rho,\gamma}^{\alpha,\beta}\left(\sum_{i=1}^{N} A_i\right)} \notag
=&\frac{1}{\sqrt{2}}  \left \| \rho^\frac{1-\alpha-\beta}{2}\left [(1-\gamma)\rho^\alpha+\gamma\rho^\beta,\sum_{i=1}^{N} A_{i} \right ] \right\|  \\  \notag
\leq&\frac{1}{\sqrt{2}}\sum_{i=1}^{N}\left\|\rho^\frac{1-\alpha-\beta}{2}
\left[(1-\gamma)\rho^\alpha+\gamma\rho^\beta,A_{i}\right]\right\|\\
=&\sum_{i=1}^{N}\sqrt{\mathrm{K}_{\rho,\gamma}^{\alpha,\beta}(A_{i})}~,\,\,\alpha,\beta\geq 0,~\alpha+\beta\leq 1,~0\leq \gamma \leq 1,
\end{align*}
and
\begin{align*}
\sqrt{\mathrm{K}_{\rho,\gamma}^{\alpha,\beta}\left(\sum_{i=1}^{N-1}A_i-A_N\right)} \notag
=&\frac{1}{\sqrt{2}}  \left \|\rho^\frac{1-\alpha-\beta}{2} \left [(1-\gamma)\rho^\alpha+\gamma\rho^\beta,\sum_{i=1}^{N-1}A_{i}-A_N \right ] \right\|  \\  \notag
\leq&\frac{1}{\sqrt{2}}\sum_{i=1}^{N}\left\|\rho^\frac{1-\alpha-\beta}{2}\left[(1-\gamma)\rho^\alpha+\gamma\rho^\beta,A_{i}\right]\right\|\\
=&\sum_{i=1}^{N}\sqrt{\mathrm{K}_{\rho,\gamma}^{\alpha,\beta}(A_{i})}~,\,\,\alpha,\beta\geq 0,~\alpha+\beta\leq 1,~0\leq \gamma \leq 1. \Box
\end{align*}

Setting $N=2$ in Theorem 2, we have the following Corollary 2.\\\hspace*{\fill}\\
{\bf Corollary 2} For arbitrary two noncommutative observables $A$
and $B$, we have
\begin{equation}\label{eq14}
\sqrt{\mathrm{K}_{\rho,\gamma}^{\alpha,\beta}(A)}+\sqrt{\mathrm{K}_{\rho,\gamma}^{\alpha,\beta}(B)}\geq \mathrm{max}\left\{\sqrt{\mathrm{K}_{\rho,\gamma}^{\alpha,\beta}(A+B)},\sqrt{\mathrm{K}_{\rho,\gamma}^{\alpha,\beta}(A-B)}\right\}~,
\end{equation}
where $\alpha,\beta\geq 0,~\alpha+\beta\leq 1,~0\leq \gamma \leq 1$.

Note that (\ref{eq14}) of Corollary 2 reduces to
(\ref{eq5}) of Theorem 2 in \cite{CB3} when
$\alpha=\beta=\frac{1}{2}$.
\\\hspace*{\fill}\\
{\bf Theorem 3} For arbitrary $N$ mutually noncommutative
observables $A_1, A_2,\cdots,A_N$ ($N>2$), we have
\begin{eqnarray}\label{eq15}
\sum_{i=1}^{N} \mathrm{K}_{\rho,\gamma}^{\alpha,\beta}(A_i)
&\geq&\frac{1}{N-2}\left[\sum_{1\leq i<j\leq N} \mathrm{K}_{\rho,\gamma}^{\alpha,\beta}(A_i+A_j)-\frac{1}{(N-1)^2}
\right.
\nonumber\\
&&\left. \left(\sum_{1\leq i<j\leq N}\sqrt{\mathrm{K}_{\rho,\gamma}^{\alpha,\beta}(A_i+A_j)}\right)^2\right],
\end{eqnarray}where $\alpha,\beta\geq 0,~\alpha+\beta\leq 1,~0\leq \gamma \leq 1$.\\\hspace*{\fill}\\
\textit{Proof} Employing the following inequality \cite{CB2},
\begin{align*}
\sum_{i=1}^{N} \| u_i\|^2
\geq&  \frac{1}{N-2}\left[\sum_{1\leq i<j\leq N} \| u_i+u_j\|^2-\frac{1}{(N-1)^2}
\left(\sum_{1\leq i<j\leq N}\| u_i+u_j\|\right)^2\right]\\ \notag
\geq& \frac{1}{2(N-1)}\sum_{1\leq i<j\leq N} \| u_i+u_j\|^2,
\end{align*}
and setting
$u_i=\rho^\frac{1-\alpha-\beta}{2}\left[(1-\gamma)\rho^\alpha+\gamma\rho^\beta,A_i\right]$
and
$u_j=\rho^\frac{1-\alpha-\beta}{2}\left[(1-\gamma)\rho^\alpha+\gamma\rho^\beta,A_j\right]$,
we obtain (\ref{eq15}). $\Box$

Note that (\ref{eq15}) of Theorem 3 reduces to
(\ref{eq12}) of Theorem 4 in \cite{CB3} when
$\alpha=\beta=\frac{1}{2}$.
\\\hspace*{\fill}\\
{\bf Theorem 4} For arbitrary $N$ mutually noncommutative
observables $A_1, A_2, \cdots, A_N$ ($N>2$), we have
\begin{eqnarray}\label{eq16}
\sum_{i=1}^{N}\sqrt{\mathrm{K}_{\rho,\gamma}^{\alpha,\beta}(A_i)}\geq \frac{1}{N-2}\left[\sum_{1\leq i<j\leq  N}\sqrt{\mathrm{K}_{\rho,\gamma}^{\alpha,\beta}(A_i+A_j)}
\right.
\nonumber\\
\left.-\sqrt{\mathrm{K}_{\rho,\gamma}^{\alpha,\beta} \left(\sum_{i=1}^{N} A_i\right)}\right]~,\,\,\alpha,\beta\geq 0,~\alpha+\beta\leq 1,~0\leq \gamma \leq 1.
\end{eqnarray}\\
\textit{Proof} By using the following inequality \cite{CB2,HRAJ,HAOY},
\begin{align*}\label{12}
\sum_{i=1}^{N} \| u_i\| \notag
\geq&  \frac{1}{N-2}\left(\sum_{1\leq i<j\leq N} \| u_i+u_j\|-
\left\| \sum_{i=1}^{N}u_i\right\| \right)\\
\geq& \left\|\sum_{i=1}^{N} u_i\right\|,
\end{align*}
with $u_i=\rho^\frac{1-\alpha-\beta}{2}\left[(1-\gamma)\rho^\alpha+\gamma\rho^\beta,A_i\right]$ and $u_j=\rho^\frac{1-\alpha-\beta}{2}\left[(1-\gamma)\rho^\alpha+\gamma\rho^\beta,A_j\right]$, we obtain (\ref{eq16}). $\Box$

Note that (\ref{eq16}) of Theorem 4 reduces to (\ref{eq14}) of Theorem 5 in \cite{CB3} when $\alpha=\beta=\frac{1}{2}$. Moreover, from the proof of Theorem 4, it can be seen that the right hand side of (\ref{eq16}) is tighter than the right hand side of (\ref{eq12}). \\\hspace*{\fill}\\
{\bf Theorem 5} For arbitrary $N$ mutually noncommutative
observables $A_1, A_2, \cdots, A_N$ ($N\geq2$), we have
\begin{eqnarray}\label{eq17}
\sum_{i=1}^{N}{\mathrm{K}_{\rho,\gamma}^{\alpha,\beta}(A_i)}\geq \frac{1}{N}\mathrm{K}_{\rho,\gamma}^{\alpha,\beta}\left(\sum_{i=1}^ {N}A_i\right)+\frac{2}{N^2(N-1)}
\nonumber\\
\left(\sum_{1\leq i <j\leq N} \sqrt{\mathrm{K}_{\rho,\gamma}^{\alpha,\beta}(A_i-A_j)}\right)^2~,
\end{eqnarray}
where $\alpha,\beta\geq 0,~\alpha+\beta\leq 1,~0\leq \gamma \leq 1$.\\\hspace*{\fill}\\
\textit{Proof} According to the lemma in \cite{ZL}, we have
\begin{align*}
\frac{2}{N^2(N-1)}\left(\sum_{1\leq i <j\leq N}\| u_i-u_j\|\right)^2
\leq& \frac{1}{N}\sum_{1\leq i <j\leq N}\| u_i-u_j\|^2\\ \notag
=&\sum_{i=1}^{N}\| u_i\|^2-\frac{1}{N}\left\|\sum_{i=1}^{N}u_i\right\|^2,
\end{align*}
that is,
\begin{align*}
\sum_{i=1}^{N}\|u_i\|^2\geq\frac{1}{N}\left\|\sum_{i=1}^{N}u_i\right\|^2
+&\frac{2}{N^2(N-1)}\left(\sum_{1\leq i <j\leq
N}\|u_i-u_j\|\right)^2.
\end{align*}
By substituting $u_i$ and $u_j$ with
$\rho^\frac{1-\alpha-\beta}{2}\left[(1-\gamma)\rho^\alpha+\gamma\rho^\beta,A_i\right]$
and $\rho^\frac{1-\alpha-\beta}{2}\left[(1-\gamma)\rho^\alpha+
\right.$\\ $\left.\gamma\rho^\beta,A_j\right]$, respectively, we
obtain (\ref{eq17}). $\Box$

Note that (\ref{eq17}) of Theorem 5 reduces to (\ref{eq12}) of Theorem 1 in \cite{ZL} when $\alpha=\beta=\frac{1}{2}$. \\\hspace*{\fill}\\
{\bf Theorem 6} For arbitrary $N$ mutually noncommutative
observables $A_1, A_2, \cdots, A_N$ ($N\geq2$), we have
\begin{eqnarray}\label{eq18}
\sum_{i=1}^{N}{\mathrm{K}_{\rho,\gamma}^{\alpha,\beta}(A_i)}
&\geq& \frac{1}{2(N-1)}\left[\frac{2}{N(N-1)}\left(\sum_{1\leq i<j\leq N} \sqrt{\mathrm{K}_{\rho,\gamma}^{\alpha,\beta}(A_i+A_j)}\right)^2 \right.
\nonumber\\
&&\left.+\sum_{1\leq i<j\leq N} \mathrm{K}_{\rho,\gamma}^{\alpha,\beta}(A_i-A_j)\right]~,\,\,\alpha,\beta\geq 0,~\alpha+\beta\leq 1,~0\leq \gamma \leq 1,
\end{eqnarray}
and
\begin{eqnarray}\label{eq19}
\sum_{i=1}^{N}{\mathrm{K}_{\rho,\gamma}^{\alpha,\beta}(A_i)}
&\geq& \frac{1}{2(N-1)}\left[\frac{2}{N(N-1)}\left(\sum_{1\leq i<j\leq N} \sqrt{\mathrm{K}_{\rho,\gamma}^{\alpha,\beta}(A_i-A_j)}\right)^2 \right.
\nonumber\\
&&\left.+\sum_{1\leq i<j\leq N} \mathrm{K}_{\rho,\gamma}^{\alpha,\beta}(A_i+A_j)\right]~,\,\,\alpha,\beta\geq 0,~\alpha+\beta\leq 1,~0\leq \gamma \leq 1.
\end{eqnarray}\\
\textit{Proof} By using the following equality,
\begin{align*}
  2(N-1)\sum_{i=1}^{N}\|u_i\|^2=\sum_{1\leq i <j\leq N}\|u_i-u_j\|^{2}+\sum_{1\leq i <j\leq N}\| u_i+u_j\|^{2}
\end{align*}
and the Cauchy-Schwarz inequality, we obtain
\begin{align*}
\sum_{1\leq i <j\leq N}\|u_i+u_j\|^{2}\geq\frac{2}{N(N-1)}\left(\sum_{1\leq i <j\leq N}\| u_i+u_j\|\right)^2,
\end{align*}
and
\begin{align*}
 \sum_{1\leq i <j\leq N}\|u_i-u_j\|^{2}\geq\frac{2}{N(N-1)}\left(\sum_{1\leq i<j\leq N}\|
 u_i-u_j\|\right)^2,
\end{align*}
respectively. Therefore, we have
\begin{align*}
\sum_{i=1}^{N}\|u_i\|^2\geq \frac{1}{2(N-1)}\left[\frac{2}{N(N-1)}\left(\sum_{1\leq i<j\leq N}\| u_i\pm u_j\|\right)^2+\sum_{1\leq i <j\leq N}\|u_i\mp u_j\|^2\right].
\end{align*}
The inequalities (\ref{eq18}) and (\ref{eq19}) follow by
replacing $u_i$ and $u_j$ with
$\rho^\frac{1-\alpha-\beta}{2}\left[(1-\gamma)\rho^\alpha+ \right.$\\
$\left.\gamma\rho^\beta,A_i\right]$ and
$\rho^\frac{1-\alpha-\beta}{2}\left[(1-\gamma)\rho^\alpha+\gamma\rho^\beta,A_j\right]$,
respectively. $\Box$

As a special case, when $\alpha=\beta=\frac{1}{2}$,
(\ref{eq18}) and (\ref{eq19}) of Theorem 6 reduce to (\ref{eq12})
and (\ref{eq13}) of Theorem 2 in \cite{ZQF}, respectively. Note also
that Theorem 5 and Theorem 6 are identical when $N=2$.
\\\hspace*{\fill}\\
{\bf Theorem 7} For arbitrary $N$ mutually noncommutative
observables $A_1, A_2, \cdots, A_N$ ($N\geq2$) and a quantum state
$\rho$, let $G$ be an $N\times N$ matrix with entries
$G_{jk}=\mathrm{Tr}(X_jX_k^\dag)$, where
$X_j=i\rho^\frac{1-\alpha-\beta}{2}[(1-\gamma)\rho^\alpha+\gamma\rho^\beta,A_j]
/\|\rho^\frac{1-\alpha-\beta}{2}[(1-\gamma)\rho^\alpha+\gamma\rho^\beta,A_j]\|$,
$i=\sqrt{-1}$. We have
\begin{equation}\label{eq20}
\sum_{j=1}^{N}\mathrm{K}_{\rho,\gamma}^{\alpha,\beta}(A_j)\geq \frac{1}{\lambda_{max}(G)}\mathrm{K}_{\rho,\gamma}^{\alpha,\beta}
\left(\sum_{j=1}^{N}A_j\right)~,\,\,\alpha,\beta\geq 0,~\alpha+\beta\leq 1,~0\leq \gamma \leq 1,
\end{equation}
where $\lambda_{max}(G)$ denotes the maximal eigenvalue of $G$.\\\hspace*{\fill}\\
\textit{Proof} It is obvious that $G$ is a positive semi-definite
matrix. Noting that
 \begin{align}
&\mathrm{K}_{\rho,\gamma}^{\alpha,\beta}\left(\sum_{j=1}^{N} A_j\right)\nonumber \\ \notag
=&-\frac{1}{2}\mathrm{Tr}\left(\left[(1-\gamma)\rho^{\alpha}
+\gamma\rho^{\beta},\sum_{j=1}^{N}A_{j}\right]^{2}\rho^{1-\alpha-\beta}\right)\\ \notag
=&\frac{1}{2}\sum_{j,k} \mathrm{Tr}\left[\left(i\rho^\frac{1-\alpha-\beta}{2}\left[(1-\gamma)\rho^\alpha
+\gamma\rho^\beta,A_{j}\right]\right)\left(i\left[(1-\gamma)\rho^\alpha
+\gamma\rho^\beta,A_{k}\right]\rho^\frac{1-\alpha-\beta}{2}\right)\right]\\ \notag
=&\frac{1}{2}\sum_{j,k }\left\|\rho^\frac{1-\alpha-\beta}{2}\left[(1-\gamma)\rho^\alpha+\gamma\rho^\beta, A_j\right]\right\| G_{jk}\left\|\rho^\frac{1-\alpha-\beta}{2}\left[(1-\gamma)\rho^\alpha+\gamma\rho^\beta, A_k\right]\right\|\\ \notag
=&\sum_{j,k}\sqrt{\mathrm{K}_{\rho,\gamma}^{\alpha,\beta}(A_{j})}G_{jk}
\sqrt{\mathrm{K}_{\rho,\gamma}^{\alpha,\beta}(A_{k})}\\ \notag
\leq &\lambda_\mathrm{max}(G)\sum_{j=1}^{N}\mathrm{K}_{\rho,\gamma}^{\alpha,\beta}(A_{j}),
\end{align}
we prove that (\ref{eq20}) holds. $\Box$

In particular, the Theorem 6 in \cite{CB3} is a special
case of our Theorem 7 when $\alpha=\beta=\frac{1}{2}$.

When $\alpha=\beta=\frac{1}{2}$, the $(\alpha,\beta,\gamma)$ WWYD skew information reduces to the WY skew information. We next compare our uncertainty relations with the existing ones. For $\alpha=\beta=\frac{1}{2}$, we denote by $LB_0, LB_1, LB_2$ and $LB_3$ the right hand sides of (\ref{eq15}), (\ref{eq17}), (\ref{eq18}) and (\ref{eq19}), respectively.\\\hspace*{\fill}\\
{\bf Example 1} Given a qubit state
$\rho=\frac{1}{2}(\mathbf{1}+\mathbf{r}\cdot\bm{\sigma})$, where
$\mathbf{r}=(x,y,z)$ is the Bloch vector satisfying
$|\mathbf{r}|\leq 1$, $\bm{\sigma}=(\sigma_1,\sigma_2,\sigma_3)$
with $\sigma_j$ $(j=1,2,3)$ the Pauli matrices, and
$\mathbf{r}\cdot\bm{\sigma}=\sum^3_{j=1}r_j\sigma_j$. The eigenvalues of $\rho$ are
$\lambda_{1,2}=(1\mp \sqrt{t})/2$, where $t=|\mathbf{r}|^2$.\\
The sum of the skew information of three Pauli operators is given by
\begin{equation}\label{eq21}
\mathrm{K}_{\rho,\gamma}^{\frac{1}{2},\frac{1}{2}}(\sigma_1)
+\mathrm{K}_{\rho,\gamma}^{\frac{1}{2},\frac{1}{2}}(\sigma_2)
+\mathrm{K}_{\rho,\gamma}^{\frac{1}{2},\frac{1}{2}}(\sigma_3)=2(1-\sqrt{1-t}).
\end{equation}
From (\ref{eq18}) and (\ref{eq15}) we have, respectively,
\begin{equation}\label{eq22}
\mathrm{K}_{\rho,\gamma}^{\frac{1}{2},\frac{1}{2}}(\sigma_1)
+\mathrm{K}_{\rho,\gamma}^{\frac{1}{2},\frac{1}{2}}(\sigma_2)
+\mathrm{K}_{\rho,\gamma}^{\frac{1}{2},\frac{1}{2}}(\sigma_3)
\geq(1-\sqrt{1-t})\left(1+\frac{xy+xz+yz}{2t}\right)+\frac{1}{12}\beta^2
\end{equation}
and
\begin{equation}\label{eq23}
\mathrm{K}_{\rho,\gamma}^{\frac{1}{2},\frac{1}{2}}(\sigma_1)
+\mathrm{K}_{\rho,\gamma}^{\frac{1}{2},\frac{1}{2}}(\sigma_2)
+\mathrm{K}_{\rho,\gamma}^{\frac{1}{2},\frac{1}{2}}(\sigma_3)\geq (1-\sqrt{1-t})\left(4-\frac{2(xy+xz+yz)}{t}\right)-\frac{1}{4}\beta^2,
\end{equation}
where
\begin{equation}\label{eq26}
\alpha=\sqrt{1-\sqrt{1-t}}\left(\sqrt{1+\frac{z^2+2xy}{t}}
+\sqrt{1+\frac{y^2+2xz}{t}}+\sqrt{1+\frac{x^2+2yz}{t}}\right).
\end{equation}
(\ref{eq19}) and (\ref{eq17}) give rise to
\begin{equation}\label{eq24}
\mathrm{K}_{\rho,\gamma}^{\frac{1}{2},\frac{1}{2}}(\sigma_1)
+\mathrm{K}_{\rho,\gamma}^{\frac{1}{2},\frac{1}{2}}(\sigma_2)
+\mathrm{K}_{\rho,\gamma}^{\frac{1}{2},\frac{1}{2}}(\sigma_3)
\geq(1-\sqrt{1-t})\left(1-\frac{xy+xz+yz}{2t}\right)+\frac{1}{12}\alpha^2
\end{equation}
and
\begin{equation}\label{eq25}
\mathrm{K}_{\rho,\gamma}^{\frac{1}{2},\frac{1}{2}}(\sigma_1)+\mathrm{K}_{\rho,\gamma}^{\frac{1}{2},\frac{1}{2}}(\sigma_2)+\mathrm{K}_{\rho,\gamma}^{\frac{1}{2},\frac{1}{2}}(\sigma_3)\geq \frac{2}{3}(1-\sqrt{1-t})\left(1-\frac{xy+xz+yz}{t}\right)+\frac{1}{9}\alpha^2,
\end{equation}
respectively, where
\begin{equation}\label{eq27}
\beta=\sqrt{1-\sqrt{1-t}}\left(\sqrt{1+\frac{z^2-2xy}{t}}
+\sqrt{1+\frac{y^2-2xz}{t}}+\sqrt{1+\frac{x^2-2yz}{t}}\right).
\end{equation}
Comparing the lower bound $LB_2$ ($LB_3$) on the right hand of the
inequality (\ref{eq18}) ((\ref{eq19})) with the bound $LB_0$ ($LB_1$)
on the right hand of inequality (\ref{eq15}) ((\ref{eq17})), we
obtain $LB_2-LB_0=(1-\sqrt{1-t})\gamma$ and
$LB_3-LB_1=(1-\sqrt{1-t})\gamma_1$, respectively, where
\begin{equation}\label{eq28}
 \gamma=-3+\frac{5(xy+xz+yz)}{2t}+\frac{1}{3}
 \left(\sqrt{1+\frac{z^2-2xy}{t}}+\sqrt{1+\frac{y^2-2xz}{t}}
 +\sqrt{1+\frac{x^2-2yz}{t}}\right)^2
\end{equation}
and
\begin{equation}\label{eq29}
\gamma_1=\frac{1}{3}+\frac{xy+xz+yz}{6t}-\frac{1}{36}\left(\sqrt{1+\frac{z^2+2xy}{t}}+\sqrt{1+\frac{y^2+2xz}{t}}+\sqrt{1+\frac{x^2+2yz}{t}}\right)^2.
\end{equation}
Note that $1-\sqrt{1-t}>0$. Denote
$x=\sqrt{t}\sin\theta\cos\varphi$,
$y=\sqrt{t}\sin\theta\sin\varphi$ and $z=\sqrt{t}\cos\theta$ with
$\theta\in[0,\pi]$ and $\varphi\in[0,2\pi]$. We have
\begin{eqnarray}\label{eq30}
\gamma
&=&
-3+\frac{5}{2}(\sin^2\theta\sin\varphi\cos\varphi+\sin\theta\cos\theta\cos\varphi+\sin\theta\cos\theta\sin\varphi)
\nonumber\\
&&+\frac{1}{3}\left(\sqrt{1+\cos^2\theta-2\sin^2\theta\sin\varphi\cos\varphi}+\sqrt{1+\sin^2\theta\sin^2\varphi-2\sin\theta\cos\theta\cos\varphi}\right.
\nonumber\\
&&\left.+\sqrt{1+\sin^2\theta\cos^2\varphi-2\sin\theta\cos\theta\sin\varphi}\right)^2>0,
\end{eqnarray}
and
\begin{eqnarray}\label{eq31}
\gamma_1
&=&
\frac{1}{3}+\frac{1}{6}(\sin^2\theta\sin\varphi\cos\varphi+\sin\theta\cos\theta\cos\varphi+\sin\theta\cos\theta\sin\varphi)
\nonumber\\
&&-\frac{1}{36}\left(\sqrt{1+\cos^2\theta+2\sin^2\theta\sin\varphi\cos\varphi}+\sqrt{1+\sin^2\theta\sin^2\varphi+2\sin\theta\cos\theta\cos\varphi}\right.
\nonumber\\
&&\left.+\sqrt{1+\sin^2\theta\cos^2\varphi+2\sin\theta\cos\theta\sin\varphi} \right)^2 >0,
\end{eqnarray}
see Fig.~\ref{fig:Fig1}.
\begin{figure}[H]\centering
\subfigure[]
{\begin{minipage}[XuCong-uncertainty-2(a)]{0.46\linewidth}
\includegraphics[width=0.9\textwidth]{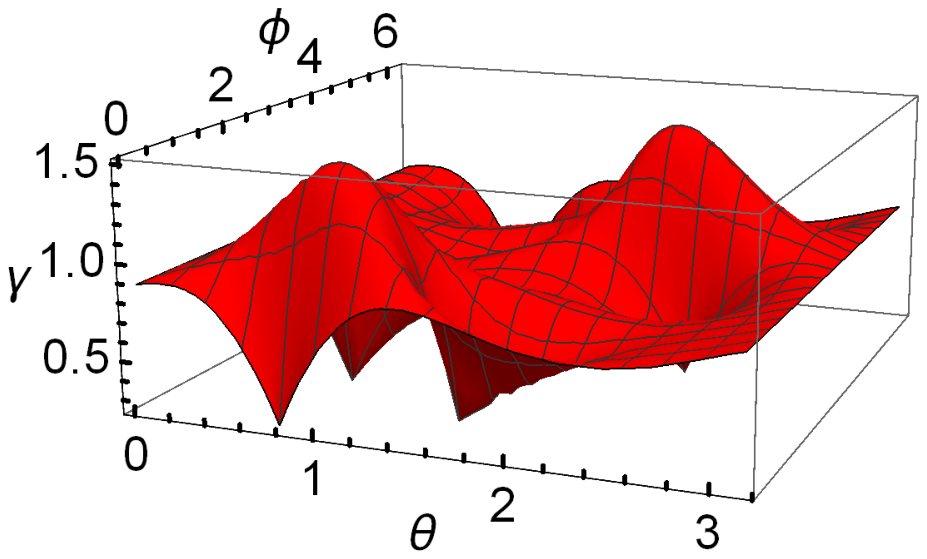}
\end{minipage}}
\subfigure[]
{\begin{minipage}[XuCong-uncertainty-2(b)]{0.46\linewidth}
\includegraphics[width=0.9\textwidth]{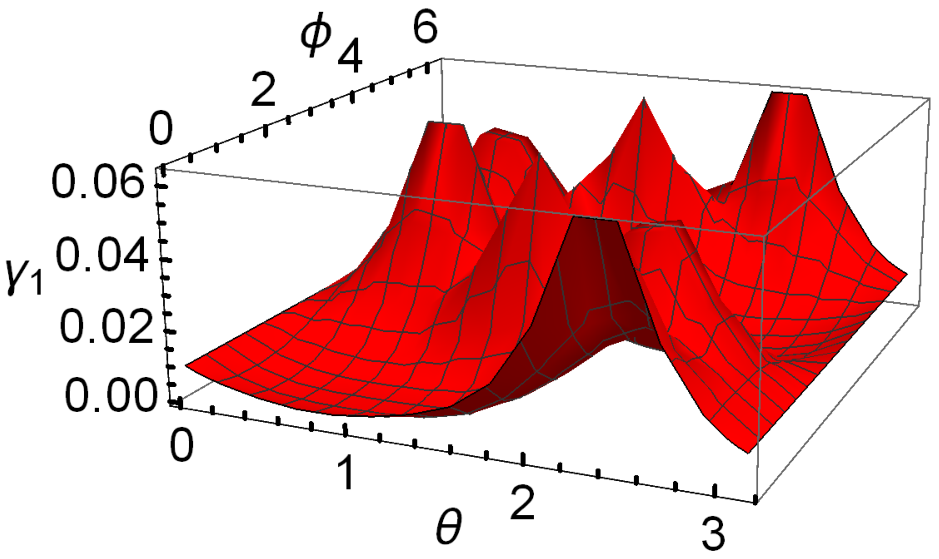}
\end{minipage}}
\caption{{$\gamma$ and $\gamma_1$ as a function of $\theta$ and
$\phi$\label{fig:Fig1}}}
\end{figure}

Figure 1 shows that our lower bound $LB_2$ ($LB_3$) is larger than
the lower bound given in \cite{CB3} (\cite{ZL}) except $t=0$ for the
case of spin-$\frac{1}{2}$. In \cite{ZQF} the authors illustrated by
an example that their lower bounds are better than the ones given in
\cite{ZL} for the case of a special qubit state with Bloch vector
$\mathbf{r}=(\frac{\sqrt{3}}{2}\cos\theta,\frac{\sqrt{3}}{2}\sin\theta,0)$.
Here, in our example we have considered the general case of
arbitrary qubit states.

Now, we give two theorems in which the lower bounds consist of the sum uncertainty
relations of different size $k$.\\\hspace*{\fill}\\
{\bf Theorem 8} For $N$ mutually noncommutative observables
$A_1,\cdots,A_N$, $2\leq k<N$, we have
\begin{eqnarray}\label{eq32}
\sum_{i=1}^{N}\mathrm{K}_{\rho,\gamma}^{\alpha,\beta}(A_i) &\geq &
\left(
                    \begin{array}{c}
                      N-2 \\
                      k-1 \\
                    \end{array}
                  \right)^{-1}
                  \left[\sum_{1\leq i_{1}<\cdots<i_{k}\leq N}\mathrm{K}_{\rho,\gamma}^{\alpha,\beta}\left(\sum_{j=1}^{k}A_{i_j}\right)-\left(
                                                                                                              \begin{array}{c}
                                                                                                                N-2 \\
                                                                                                                k-2 \\
                                                                                                              \end{array}
                                                                                                            \right)
                 \left(
                   \begin{array}{c}
                     N-1 \\
                     k-1 \\
                   \end{array}
                 \right)^{-2}
                 \right.
                 \nonumber\\
                 &&\left. \left(\sum_{1\leq i_{1}<\cdots<i_{k}\leq N}\sqrt{\mathrm{K}_{\rho,\gamma}^{\alpha,\beta}\left(\sum_{j=1}^{k}A_{i_j}\right)}\right)^{2}
                  \right]~,\alpha,\beta\geq 0,\alpha+\beta\leq 1,0\leq \gamma \leq 1.
\end{eqnarray}
\textit{Proof} It is a direct result of the following inner product
inequality proved in the Theorem 1 in \cite{CAL},
\begin{eqnarray*} \sum_{i=1}^{N}\|u_i\|^2 &\geq & {\left(
                              \begin{array}{c}
                                N-2 \\
                                k-1 \\
                              \end{array}
                            \right)}^{-1}\left[\sum_{1\leq i_{1}<\cdots<i_{k}\leq N}\|u_{i_1}+\cdots+u_{i_k}\|^2-\left(
                                                                                                            \begin{array}{c}
                                                                                                              N-2 \\
                                                                                                              k-2 \\
                                                                                                            \end{array}
                                                                                                          \right)
                           \left(
                             \begin{array}{c}
                               N-1 \\
                               k-1 \\
                             \end{array}
                           \right)^{-2}
                           \right.
                 \nonumber\\
                 &&\left.\left(\sum_{1\leq i_{1}<\cdots<i_{k}\leq N}
                 {\|u_{i_1}+\cdots+u_{i_k}\|}\right)^2\right].
\end{eqnarray*}
By substituting $u_i$ and $u_{i_j}$ ($1\leq j\leq k$) with
$\rho^\frac{1-\alpha-\beta}{2}\left[(1-\gamma)\rho^\alpha+\gamma\rho^\beta,A_i\right]$
and
$\rho^\frac{1-\alpha-\beta}{2}\left[(1-\right.$\\
$\left.\gamma)\rho^\alpha+\gamma\rho^\beta,A_{i_j}\right]$,
respectively, we obtain (\ref{eq32}). $\Box$

We observe that Theorem 3 is a special case of Theorem 8 when $k=2$. Similarly,
we can give a variation of the Theorem 4 for the sum uncertainty relation
of $\sqrt{\mathrm{K}_{\rho,\gamma}^{\alpha,\beta}(A_i)}$. \\\hspace*{\fill}\\
{\bf Theorem 9} For $N$ mutually noncommutative observables
$A_1,\cdots,A_N$, we have
\begin{eqnarray}\label{eq33}
\sum_{i=1}^{N}\sqrt{\mathrm{K}_{\rho,\gamma}^{\alpha,\beta}(A_i)}\geq
\sum_{1\leq i_{1}<\cdots<i_{N-1}\leq
N}\sqrt{\mathrm{K}_{\rho,\gamma}^{\alpha,\beta}(A_{i_1}+\cdots+A_{i_{N-1}})}
\nonumber\\
-(N-2)\sqrt{\mathrm{K}_{\rho,\gamma}^{\alpha,\beta}\left(\sum_{i=1}^{N}A_{i}\right)}~,\,\,\alpha,\beta\geq 0,~\alpha+\beta\leq 1,0\leq \gamma \leq 1.
\end{eqnarray}
\textit{Proof} It is a direct result of the following inner product inequality,
\begin{equation*}
\sum_{i=1}^{N}\|u_i\|\geq \sum_{1\leq i_{1}<\cdots<i_{N-1}\leq N}
\|u_{i_1}+\cdots+u_{i_{N-1}}\|-(N-2)\left\|\sum_{i=1}^{N}u_i\right\|,
\end{equation*}
which has been proved in the Theorem 2 in \cite{CAL}.
Replacing $u_i$ and $u_{i_j}$ ($1\leq j\leq N-1$) by
$\rho^\frac{1-\alpha-\beta}{2}\left[(1-\gamma)\rho^\alpha+\gamma\rho^\beta,A_i\right]$
and $\rho^\frac{1-\alpha-\beta}{2}\left[(1-\gamma)\rho^\alpha+\gamma\rho^\beta,A_{i_j}\right]$, respectively, we get (\ref{eq33}) immediately. $\Box$\\\hspace*{\fill}\\

From the above results we have, for arbitrary $N$
mutually noncommutative observables $A_1, A_2, \cdots, A_N$,
\begin{equation*}
 \sum_{i=1}^{N}\mathrm{K}_{\rho,\gamma}^{\alpha,\beta}(A_i)\geq \mathrm{max}\{thm 1,thm 3,thm 5,thm 6,thm 7,thm 8\};
 \end{equation*}
\begin{equation*}
 \sum_{i=1}^{N}\sqrt{\mathrm{K}_{\rho,\gamma}^{\alpha,\beta}(A_i)}\geq \mathrm{max}\{thm 2,thm 4, thm 9 \},
 \end{equation*}
where thm 1, thm 3, thm 5, thm 6, thm 7, thm 8, thm 2, thm 4 and thm
9 stand for the lower bounds in Theorem 1, Theorem 3, Theorem 5,
Theorem 6, Theorem 7, Theorem 8, Theorem 2, Theorem 4 and Theorem 9,
respectively.

\vskip0.1in

\noindent {\bf 4. Sum uncertainty relations for quantum channels in
terms of ($\alpha,\beta,\gamma$) MWWYD skew information}\\\hspace*{\fill}\\
In this section, we explore the uncertainty relations for arbitrary
$N$ quantum channels in terms of ($\alpha,\beta,\gamma$) MWWYD skew
information. By using the same techniques, it can be
seen that the results in Section 2 also hold if the observables
$A_i\in \mathcal{S(H)}$ $(i=1,2,\cdots,N)$ are replaced by $E_i\in
\mathcal{B(H)}$ $(i=1,2,\cdots,N)$ (which are not necessarily
Hermitian). Therefore, the results in this section are easy
consequences by imitating the proofs of Theorem 1, Theorem 3, Theorem
4, Theorem 5 and Theorem 6 in Section 3, Theorem 2 in \cite{FSS} and
the definition of ($\alpha,\beta,\gamma$) MWWYD skew information
with respect to quantum channels in Eq. (\ref{eq9}). Hence we only
sketch the proof of Theorem 11 and omit the proofs of the rest theorems.
\\\hspace*{\fill}\\
{\bf Theorem 10} Let $\Phi_{1},\cdots,\Phi_N$ be $N$ quantum
channels with Kraus representations
$\Phi_t(\rho)=\sum_{i=1}^{n}E_{i}^{t}\rho (E_{i}^{t})^\dag$,
$t=1,2,\cdots,N$ ($N>2$), we have
\begin{eqnarray}\label{eq34}
\sum_{t=1}^{N}\mathrm{K}_{\rho,\gamma}^{\alpha,\beta}(\Phi_t)
&\geq&\mathop{\mathrm{max}}\limits_{\pi_t,\pi_s\in S_n}\frac{1}{2(N-1)}\left\{\sum_{1\leq t<s\leq N}\sum_{i=1}^{n} \mathrm{K}_{\rho,\gamma}^{\alpha,\beta}(E_{\pi_{t}(i)}^{t}\pm E_{\pi_{s}(i)}^{s})\right\},
\end{eqnarray}
where $\alpha,\beta\geq
0,~\alpha+\beta\leq 1,~0\leq \gamma \leq 1$, $S_n$ is the n-element permutation group and $\pi_{t},\pi_{s}\in S_n$ are arbitrary n-element permutations.\\\hspace*{\fill}\\
\textit{Proof} By using the inequality in the proof of Theorem 1, we
obtain
\begin{eqnarray*}
\sum_{t=1}^{N}\mathrm{K}_{\rho,\gamma}^{\alpha,\beta}(E_{\pi_{t}(i)}^{t})
&\geq&\frac{1}{2(N-1)}\left\{\sum_{1\leq t<s\leq N} \mathrm{K}_{\rho,\gamma}^{\alpha,\beta}(E_{\pi_{t}(i)}^{t}\pm E_{\pi_{s}(i)}^{s})\right\},
\end{eqnarray*}
where $\alpha,\beta\geq 0,~\alpha+\beta\leq 1,~0\leq \gamma \leq 1$. By the definition of Eq. (\ref{eq9}), the conclusion follows immediately. $\Box$
\\\hspace*{\fill}\\
{\bf Theorem 11} Let $\Phi_{1},\cdots,\Phi_N$ be $N$ quantum
channels with Kraus representations
$\Phi_t(\rho)=\sum_{i=1}^{n}E_{i}^{t}\rho (E_{i}^{t})^\dag,
~t=1,2,\cdots,N$ ($N>2$), we have
\begin{eqnarray}\label{eq35}
\sum_{t=1}^{N}\mathrm{K}_{\rho,\gamma}^{\alpha,\beta}(\Phi_t)
&\geq& \mathop{\mathrm{max}}\limits_{\pi_t,\pi_s\in S_n}\frac{1}{N-2}\left\{\sum_{1\leq t<s\leq N}\sum_{i=1}^{n}\mathrm{K}_{\rho,\gamma}^{\alpha,\beta}(E_{\pi_{t}(i)}^{t}+E_{\pi_{s}(i)}^{s}) \right.
\nonumber\\
&&\left.-\frac{1}{(N-1)^{2}}\left[\sum_{i=1}^{n}\left(\sum_{1\leq t<s\leq N}\sqrt{\mathrm{K}_{\rho,\gamma}^{\alpha,\beta}(E_{\pi_{t}(i)}^{t}
+E_{\pi_{s}(i)}^{s})}\right)^{2}\right]\right\},
\end{eqnarray}
where $\alpha,\beta\geq
0,~\alpha+\beta\leq 1,~0\leq \gamma \leq 1$, $S_n$ is the n-element permutation group and $\pi_{t},\pi_{s}\in S_n$ are arbitrary n-element permutations.

In particular, Theorem 2 in \cite{ZL} is a special
case of our Theorem 11 with
$\alpha=\beta=\frac{1}{2}$.\\\hspace*{\fill}\\
{\bf Theorem 12} Let $\Phi_{1},\cdots,\Phi_N$ be $N$ quantum
channels with Kraus representations
$\Phi_t(\rho)=\sum_{i=1}^{n}E_{i}^{t}\rho (E_{i}^{t})^\dag,~
t=1,2,\cdots,N$ ($N>2$), we have
\begin{eqnarray}\label{eq36}
\sum_{t=1}^{N}\sqrt{\mathrm{K}_{\rho,\gamma}^{\alpha,\beta}(\Phi_t)}
&\geq& \mathop{\mathrm{max}}\limits_{\pi_t,\pi_s\in S_n}\frac{1}{N-2}
\left\{\sum_{1\leq t<s\leq N}\sum_{i=1}^{n}\sqrt{\mathrm{K}_{\rho,\gamma}^{\alpha,\beta}(E_{\pi_{t}(i)}^{t}
+E_{\pi_{s}(i)}^{s})} \right.
\nonumber\\
&&\left.-\sum_{i=1}^{n}\sqrt{\mathrm{K}_{\rho,\gamma}^{\alpha,\beta}
\left(\sum_{t=1}^{N}E_{\pi_{t}(i)}^{t} \right)} \right\},
\end{eqnarray}
where $\alpha,\beta\geq
0,~\alpha+\beta\leq 1,~0\leq \gamma \leq 1$, $S_n$ is the n-element permutation group and $\pi_{t},\pi_{s}\in S_n$ are arbitrary n-element permutations.\\\hspace*{\fill}\\
{\bf Theorem 13} Let $\Phi_{1},\cdots,\Phi_N$ be $N$ quantum
channels with Kraus representations
$\Phi_t(\rho)=\sum_{i=1}^{n}E_{i}^{t}\rho (E_{i}^{t})^\dag,
t=1,2,\cdots,N$ ($N\geq2$), we have
\begin{eqnarray}\label{eq37}
\sum_{t=1}^{N}\mathrm{K}_{\rho,\gamma}^{\alpha,\beta}(\Phi_t)
&\geq& \mathop{\mathrm{max}}\limits_{\pi_t,\pi_s\in S_n}\left\{\frac{1}{N}\sum_{i=1}^{n}\mathrm{K}_{\rho,\gamma}^{\alpha,\beta}
\left(\sum_{t=1}^{N}E_{\pi_{t}(i)}^{t}\right) \right.
\nonumber\\
&&\left.+\frac{2}{N^{2}(N-1)}\left[\sum_{i=1}^{n}\left(\sum_{1\leq t<s\leq N}\sqrt{\mathrm{K}_{\rho,\gamma}^{\alpha,\beta}(E_{\pi_{t}(i)}^{t}
-E_{\pi_{s}(i)}^{s})}\right)^{2}\right]\right\},
\end{eqnarray}
where $\alpha,\beta\geq
0,~\alpha+\beta\leq 1,~0\leq \gamma \leq 1$, $S_n$ is the n-element permutation group and $\pi_{t},\pi_{s}\in S_n$ are arbitrary n-element permutations.

In particular, when $\alpha=\beta=\frac{1}{2}$, (\ref{eq37}) of Theorem 14 reduces to (\ref{eq32}) of Theorem 3 in \cite{ZL}.\\\hspace*{\fill}\\
{\bf Theorem 14} Let $\Phi_{1},\cdots,\Phi_N$ be $N$ quantum
channels with Kraus representations
$\Phi_t(\rho)=\sum_{i=1}^{n}E_{i}^{t}\rho (E_{i}^{t})^\dag,~
t=1,2,\cdots,N$ ($N\geq2$), we have
\begin{eqnarray}\label{eq38}
\sum_{t=1}^{N}\mathrm{K}_{\rho,\gamma}^{\alpha,\beta}(\Phi_t)
&\geq& \mathop{\mathrm{max}}\limits_{\pi_t,\pi_s\in S_n}\frac{1}{2(N-1)}\left\{\frac{2}{N(N-1)}\left[\sum_{i=1}^{n}\left(\sum_{1\leq t<s\leq N}\sqrt{\mathrm{K}_{\rho,\gamma}^{\alpha,\beta}(E_{\pi_{t}(i)}^{t}
+E_{\pi_{s}(i)}^{s})}\right)^2\right] \right.
\nonumber\\
&&\left.+\sum_{1\leq t<s\leq N}\sum_{i=1}^{n}\mathrm{K}_{\rho,\gamma}^{\alpha,\beta}(E_{\pi_{t}(i)}^{t}
-E_{\pi_{s}(i)}^{s})\right\},
\end{eqnarray}
and
\begin{eqnarray}\label{eq39}
\sum_{t=1}^{N}\mathrm{K}_{\rho,\gamma}^{\alpha,\beta}(\Phi_t)
&\geq& \mathop{\mathrm{max}}\limits_{\pi_t,\pi_s\in S_n}\frac{1}{2(N-1)}\left\{\frac{2}{N(N-1)}\left[\sum_{i=1}^{n}\left(\sum_{1\leq t<s\leq N}\sqrt{\mathrm{K}_{\rho,\gamma}^{\alpha,\beta}(E_{\pi_{t}(i)}^{t}
-E_{\pi_{s}(i)}^{s})}\right)^2\right] \right.
\nonumber\\
&&\left.+\sum_{1\leq t<s\leq N}\sum_{i=1}^{n}\mathrm{K}_{\rho,\gamma}^{\alpha,\beta}(E_{\pi_{t}(i)}^{t}
+E_{\pi_{s}(i)}^{s})\right\},
\end{eqnarray}
where $\alpha,\beta\geq 0,~\alpha+\beta\leq 1,~0\leq \gamma \leq 1$, $S_n$ is the n-element permutation group and
$\pi_{t},\pi_{s}\in S_n$ are arbitrary n-element permutations.

For arbitrary $N$ matrices (not necessarily Hermitian) $E_1,\cdots,
E_N$, we call an $N\times N$ matrix $G$ a covariance matrix of
$E_1,\cdots, E_N$ if the entries of $G$ are given by
\begin{align*}
 G_{jk}=\mathrm{Tr}\frac{(i[(1-\gamma)\rho^\alpha+\gamma\rho^\beta,E_j]
 \rho^\frac{1-\alpha-\beta}{2})\cdot (i[(1-\gamma)\rho^\alpha+\gamma\rho^\beta,E_k]\rho^\frac{1-\alpha-\beta}{2})^\dag}
 {\|\rho^\frac{1-\alpha-\beta}{2}[(1-\gamma)\rho^\alpha
 +\gamma\rho^\beta,E_j]\|\cdot\|\rho^\frac{1-\alpha-\beta}{2}
 [(1-\gamma)\rho^\alpha+\gamma\rho^\beta,E_k]\|},
\end{align*}
where $\|\cdot\|$ is a norm on the $d$-dimensional complex linear space of matrices $M_d(\mathbb{C})$. It can be verified that $G$ is a positive semi-definite complex matrix. Analogizing the idea in \cite{FSS}, we obtain the following theorem.\\\hspace*{\fill}\\
{\bf Theorem 15} Let $\Phi_{1},\cdots,\Phi_N$ be $N$ quantum
channels with Kraus representations
$\Phi_t(\rho)=\sum_{i=1}^{n}E_{i}^{t}\rho (E_{i}^{t})^\dag,~
t=1,2,\cdots,N$ ($N\geq2$), we have
\begin{eqnarray}\label{eq40}
\sum_{t=1}^{N}\mathrm{K}_{\rho,\gamma}^{\alpha,\beta}(\Phi_t)
\geq\frac{1}{\lambda_{\mathrm{max}}(G)}\mathrm{K}_{\rho,\gamma}^{\alpha,\beta}
\left(\sum_{i=1}^{n}\sum_{t=1}^{N}E_{i}^{t}\right),
\end{eqnarray}
where $\alpha,\beta\geq0,~\alpha+\beta\leq 1,~0\leq \gamma \leq 1$, and $G$ is the $Nn\times Nn$ covariance matrix of $\{E_i^{t}\}_{1\leq i\leq n,1\leq t\leq N}$.

Note that (\ref{eq40}) of Theorem 15 reduces to (\ref{eq2}) of
Theorem 2 in \cite{FSS} when $\alpha=\beta=\frac{1}{2}$ and $N=2$.

\vskip0.1in

\noindent {\bf 5. Conclusions}\\\hspace*{\fill}\\
We have introduced the ($\alpha,\beta,\gamma$) weighted
Wigner-Yanase-Dyson (($\alpha,\beta,\gamma$) WWYD) skew information
and the ($\alpha,\beta,\gamma$) modified weighted
Wigner-Yanase-Dyson (($\alpha,\beta,\gamma$) MWWYD) skew
information, which are more general than the previous concepts. We
have derived sum uncertainty relations for $N$ mutually
noncommutative observables based on the $(\alpha,\beta,\gamma)$ WWYD
skew information, which includes the results in \cite{CB3} as
special cases. Following the idea in Ref. \cite{CB3,CAL}, we have
derived other forms of lower bounds using the sum of the
$(\alpha,\beta,\gamma)$ WWYD skew information for any other number
of observables less than $N$. It is found that when
$\alpha=\beta=\frac{1}{2}$, for the spin-$\frac{1}{2}$ case and the
observables of Pauli operators $\sigma_1,\sigma_2,\sigma_3$, our
lower bounds $LB_2$ and $LB_3$ are tighter than the existing ones.
Finally, we have also explored sum uncertainty relations for quantum
channels in terms of the $(\alpha,\beta,\gamma)$ MWWYD skew
information. The results in this paper cover the ones in \cite{CB3}
and \cite{ZL} for WY skew information, and may shed some new light
on the study of skew information-based sum uncertainty relations for
observables and quantum channels.

\vskip0.1in

\noindent

\subsubsection*{Acknowledgements}
\small {This work was supported by National Natural Science
Foundation of China (Grant Nos. 12161056, 11701259, 12075159,
12171044); Jiangxi Provincial Natural Science Foundation (Grant No.
20202 BAB201001); Beijing Natural Science Foundation (Grant No.
Z190005); Academy for Multidisciplinary Studies, Capital Normal
University; Shenzhen Institute for Quantum Science and Engineering,
Southern University of Science and Technology (Grant No.
SIQSE202001); the Academician Innovation Platform of Hainan
Province.}

\end{document}